\documentclass[aps,prl,twocolumn,superscriptaddress]{revtex4-2}

\usepackage[T1]{fontenc}                                
\usepackage[USenglish]{babel}							
\usepackage{newtxmath,newtxtext}						
\usepackage{soul}										

\usepackage{amsmath, amsfonts, bm}
\usepackage{mathtools}

\usepackage[colorlinks=false,hidelinks,pdfpagelabels]{hyperref}
\usepackage{cleveref}		

\usepackage{graphicx}

\begin{document}
	
	\title{Emergent order in a continuous state adaptive network model of living systems}

	\author{Carsten T. van de Kamp}

	\affiliation{Delft Institute of Applied Mathematics, Delft University of Technology, The Netherlands} 
	\affiliation{Department of Bionanoscience, Kavli Institute of Nanoscience, Delft University of Technology, The Netherlands}
	
	\author{George Dadunashvili}

	\affiliation{Department of Bionanoscience, Kavli Institute of Nanoscience, Delft University of Technology, The Netherlands}
	
	\author{Johan L.A. Dubbeldam}
	\email[]{J.L.A.Dubbeldam@TUDelft.nl}

	\affiliation{Delft Institute of Applied Mathematics, Delft University of Technology, The Netherlands}
	
	\author{Timon Idema}
	\email[]{T.Idema@TUDelft.nl}

	\affiliation{Department of Bionanoscience, Kavli Institute of Nanoscience, Delft University of Technology, The Netherlands}
	
	\date{\today}
	
	\begin{abstract}
    Order can spontaneously emerge from seemingly noisy interactions between biological agents, like a flock of birds changing their direction of flight in unison, without a leader or an external cue. We are interested in the generic conditions that lead to such emergent phenomena. To find these conditions, we use the framework of complex networks to characterize the state of agents and their mutual influence. We formulate a continuous state adaptive network model, from which we obtain the phase boundaries between swarming and disordered phases and characterize the order of the phase transition.
	\end{abstract}
	

	\maketitle
	
Swarming is a collective phenomenon which is encountered  in many biological  systems, such as schools of fish, swarms of locusts and flocks of birds. Some swarming phenomena in biological systems can be guided by external cues, or a leading individual, but a large variety of biological systems display spontaneously emerging, self-organized group behavior~\cite{vicsek_collective_2012}. Even unicellular organisms, such as the bacterium \emph{E. coli} and the amoeba \emph{D. discoideum}, have been observed to display such collective behavior \cite{bouffanais_design_2016}.

Models of swarm formation can be broadly put into two categories. The first group we call mechanistic models. Their microscopic interactions are solely based on first principles, and they do not encode the tendency to order explicitly. This works well if all microscopic forces that determine the dynamics are known. Such models have been successfully used to describe the dynamics of self propelled particles of different shapes that exhibit exclusion interactions~\cite{van_drongelen_collective_2015, mccusker_active_2019, echten_defect_2020}. 
The second group, that we call heuristic models, are well exemplified by opinion formation in social groups. We know that people can convince others to join their cause. However, we do not yet understand the details of how one individual can cause another to form an opinion, thus we have a need for a heuristic rule that postulates how two individuals can form a consensus. Such heuristic treatment is in general needed when it is clear that there is a microscopic tendency to align, but we do not know the details of interactions; in context of opinion dynamics, such a model was first put forward by Vicsek et al.~\cite{vicsek_novel_1995}. 

\begin{figure}[t!]
    \centering
    \includegraphics{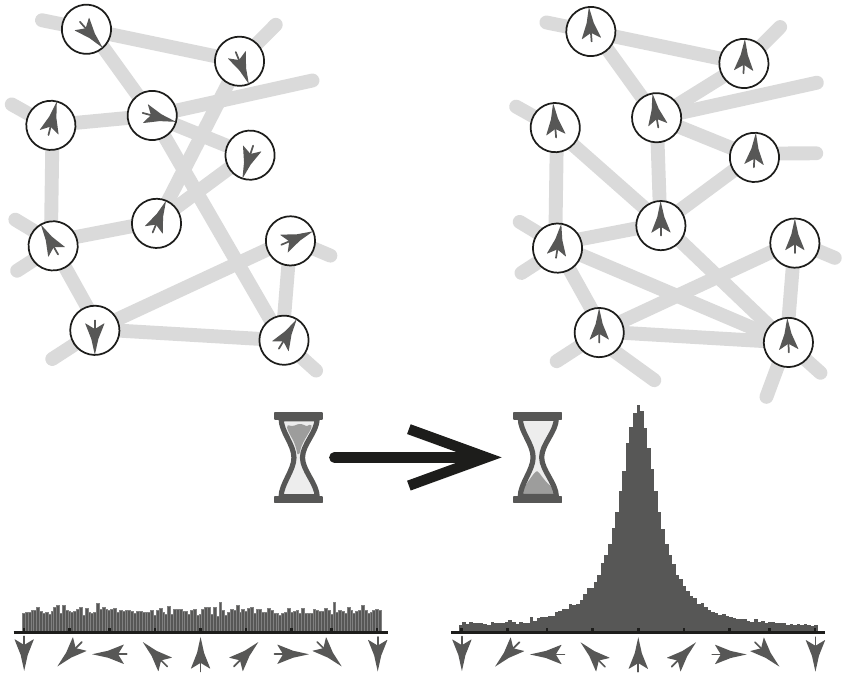}
    \caption{An illustration of a swarming phase transition in an adaptive network.
    \textbf{Top:} Each node represents an agent that can point in any direction on a plane. These agents are connected to each other through links, which represent mutual awareness. \textbf{Top left:} The agents do not have a common heading direction and the system is disordered. \textbf{Top right:}  Through interactions the agents can form a coherently moving swarm and choose an overall heading direction.
    \textbf{Bottom:} How often a certain state occurs in the network tells us if the network is ordered or disordered. \textbf{Bottom left:} different states occur with roughly the same frequency, whereas on the \textbf{Bottom right:} one direction is the most common.}
    \label{fig:graphical_abstract}
\end{figure}

Giving up the connection to the mechanistic first principles opens a door to an interesting opportunity. One can recognize that the heuristic of \emph{be more like your neighbor} is now the central aspect of the evolution of the system, instead of spatial dynamics of the particles. Thus we are shifting our attention from spatial dynamics to the topology of interactions, a problem for which the language of complex networks is uniquely well suited. In this framework the system is represented by a dynamic network. The agents, corresponding to the nodes of the network, can occupy different states, representing the direction of movement, and the changing topology of the network encodes the temporal dynamics of the interactions. An example of such a system can be seen in \cref{fig:graphical_abstract}. One such dynamic network model, for a system with a discrete state space, has been proposed by Chen, Huepe and Gross \cite{chen_adaptive_2016}. This model predicts a first or second order phase transition from a disordered to an ordered state, depending on the number of possible internal states. The first order phase transition has previously been observed in agent based models that follow the spatial dynamics of individuals, and has been confirmed both in experiments with active matter and observations of living systems~\cite{buhl_disorder_2006, makris_critical_2009, huber_emergence_2018}. The character of the transition can also depend on finite system size effects~\cite{gregoire_onset_2004, nagy_new_2007} and on subtle changes in the way the noise is incorporated~\cite{aldana_phase_2007}.
Earlier network models \cite{sood_voter_2005,huepe_adaptive-network_2011, chen_adaptive_2016} have focused on a discrete state space. All the aforementioned biological systems, however, have a continuous state space. Our intuition from equilibrium statistical mechanics tells us that the nature of the state space can matter a lot~\cite{onsager_crystal_1944, mermin_absence_1966, merkl_recurrent_1994}. It has been demonstrated that in nonequilibrium systems, details affect the type of phase transition. Therefore, in this paper we investigate the swarming transition in a continuous state space.

In this letter we present the numerical solutions of our model accompanied with analytic expressions for the phase space boundary, that we found through stability analysis of the disordered phase. Our results clearly show that continuous state adaptive network models are able to describe spontaneous emergence of order in active systems. We find that the balance of time scales in the model determines the type of phase transition. Our model has two main time scales, the time scale of information propagation through the network, and the time scale of network reshaping. When the latter is significantly faster than the former, we get a mean field like situation, where the network is updated  so fast that every node has contact with a random selection of nodes, replacing the link structure by an average connectivity number. The phase transition in this case becomes second order, while in the case that the time scales are of comparable magnitude, we find a first order transition. A meta phase transition between the first and second order regimes, similar to this one has been reported before in the context of active adaptive systems~\cite{nagy_new_2007}.

\emph{\label{sec:model}Model -- }
We model a system of self-propelled particles with a constant speed and changing direction in two dimensions. Each particle corresponds to a node in a network, with an internal state that represents the direction of movement. We represent this internal state by an angle $\theta \in (-\pi,\pi]$. Nodes may be connected by links, indicating mutual awareness. We will refer to individuals connected by a link as neighbors.

In our system not only the states of individual agents can evolve dynamically, but also the relationship of the agents with their surroundings. Analogously to \cite{chen_adaptive_2016}, we distinguish four types of dynamics, given by four rules: \\
\textbf{Rule~1} Individuals spontaneously change their heading direction to another uniformly chosen direction with rate $w_0$. \\
\textbf{Rule~2} Individuals adopt to the average direction of two neighbors with rate $w_2$. \\
\textbf{Rule~3} Arbitrarily chosen not-neighboring individuals become neighbors with a coupling rate $c$. \\
\textbf{Rule~4} Arbitrarily chosen neighbors loose mutual awareness with a decoupling rate $d$.

\begin{figure}[t]
	\includegraphics{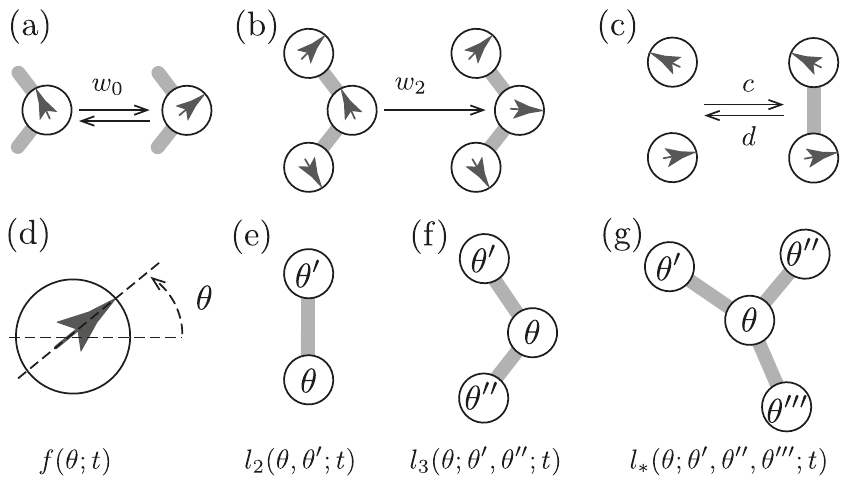}
	\caption{
	\textbf{(a-c)} Illustration of the dynamic rules of the model. The internal state of each node (circle) is represented by the direction of the arrow. These dynamics take place irrespective of any additional links that may be present, but are not drawn.
	\textbf{(d-g)} Visualization of interaction subgraphs. Every subgraph depicts a node in state $\theta$ that is interacting with its neighbors.
	$f(\theta)$, $l_2(\theta, \theta';t)$, $l_3(\theta; \theta', \theta'';t)$ and $l_*(\theta;\theta',\theta''\theta''';t)$ are the relative densities of the respective subgraphs in the network at time $t$.}
	\label{fig:interaction_subgraphs}
\end{figure}

An illustration of the model is given in \cref{fig:interaction_subgraphs}. 
The first two rules are comparable to the rules in the Vicsek model \cite{vicsek_novel_1995}; rule~1 is similar to the noise that is added to each particle's updated heading direction, whereas rule~2 corresponds to the tendency of individuals to align with their neighbors within a certain radius. The radius is modeled in the network with the links, since we do not keep track of the physical position of individuals in space. The interactions described by rule~3 and rule~4 are needed since non-neighboring individuals, moving in different directions, might become aware of each other and conversely individuals which are neighboring, but head in different directions, may loose mutual awareness; therefore, their link must be created or removed respectively. Since we want to look at the behavior of groups comprised of agents of the same species, we assume that the rates are global.
We define the state distribution function $f(\theta;t)$ which represents the fraction of the agents in the state $\theta \in (-\pi,\pi]$ at time $t$. Thus the integral of $f(\theta,t)$ over the whole domain is normalized at any time. The cartoon in \cref{fig:graphical_abstract} is an illustration of such a state density that starts out as a random, noisy distribution over the states and evolves into a state corresponding to a collective motion in one predominant direction.
Moreover, we define the link distribution function $l_2(\theta, \theta';t)$ as the density of neighboring individuals, one of which is heading in direction $\theta$ while the other one has direction $\theta'$, at time $t$. Note that in contrast to $f(\theta;t)$, the link density $l_2(\theta, \theta';t)$ is not normalized since the overall number of links can grow or shrink, making the normalization of $l_2$ time dependent. We also define higher order interaction terms like the path-like three node and star-like four node subgraph densities $l_3(\theta;\theta',\theta'';t)$ and $l_*(\theta;\theta',\theta'',\theta''';t)$ analogously. A visual representation of these subgraphs can be seen in \cref{fig:interaction_subgraphs}.
We note that the link density functions obey certain symmetries. In particular, we have $l_2(\theta,\theta';t) = l_2(\theta',\theta;t)$ and for higher order terms all non central nodes can be exchanged in any order.

One can derive the master equations for the continuous state model using the four rules, which govern the dynamics of the state and link density distributions. These equations can be interpreted as the extension of the master equations in \cite{chen_adaptive_2016}, into  the continuous state set.
\begin{subequations}
	\label{eq:cont_complete}
	\begin{alignat}{2}
	\partial_t f(\theta;t)
	=&\ w_0 \left( \frac{1}{2\pi} - f(\theta;t) \right) 
	+  w_2 F^{\mathrm{int}}\left[l_3;\theta;t\right]\label{eq:cont_f} \\
	\partial_t l_2(\theta,\theta';t)
	=&\
	w_0 L^{\mathrm{noise}}\left[l_2;\theta,\theta';t \right] + w_2 L^{\mathrm{int}}\left[l_3, l_*;\theta,\theta';t \right]  \nonumber \\
	&\ + c \ f(\theta;t)\ f(\theta';t) -  d \ l_2(\theta,\theta';t)\label{eq:cont_l} .
	\end{alignat}
\end{subequations}
The first term of \cref{eq:cont_f} scales with $w_0$ and models the spontaneous direction changes of nodes. $F^{\mathrm{int}}$ in the second term is a functional that describes three-body interactions that cause a change of the state of an agent. The \cref{eq:cont_l} contains four contributions. $L^{\mathrm{noise}}$ accounts for changes in link density, due to the random changes of states of already linked agents. $L^{\mathrm{int}}$ represents the changes of the link density due to the changes of states of already linked agents, but in this case the changes were induced by interactions between neighbors. The last two terms represent the change in link density due to link creation or deletion and scale with the rates $c$ and $d$ respectively. 
Both $L^{\mathrm{noise}}$ and $L^{\mathrm{int}}$ integrate to zero over the whole domain~\cite{supplement} i.e. $\int_{-\pi}^{\pi}\mathrm{d}\theta\; \int_{-\pi}^{\pi}\mathrm{d}\theta'\ L=0$, thus we can show that the overall link density, that we define as
\begin{equation}
    k(t)=\int_{-\pi}^{\pi}\mathrm{d}\theta\; \int_{-\pi}^{\pi}\mathrm{d}\theta'\ l_2(\theta,\theta';t),
\end{equation}
always evolves to a steady state. Integrating \cref{eq:cont_l} over $\theta$ and $\theta'$ yields
a simple differential equation $\partial_{t} k(t) = c - d k(t)$ which has the solution
\begin{equation}\label{eq:overall_link_density}
    k(t) = \left(k(0)-\frac{c}{d}\right)\exp(-dt) + \frac{c}{d}.
\end{equation}
Thus we find for the steady state link density $k_s=c/d$. During our analysis we found that the system was not qualitatively affected by changes in the value of $k_s$~\cite{supplement}, so from now on, unless explicitly stated otherwise, we will assume that $k_s=1$, i.e. $c=d$. We will refer to both rates as the rate of link dynamics.
The parameter space can be reduced further by rescaling time with the rate $w_2$. Going forward we will set $w_2=1$, which implies that time is measured in units of $1/w_2$ and all other rates are measured in multiples of $w_2$.

In order to solve \cref{eq:cont_complete} we need to relate higher order subgraph density functions $l_3$ and $l_*$ to the link density function $l_2$
\begin{align}
l_3(\theta;\theta',\theta'';t) & = \frac{l_2(\theta,\theta';t)\ l_2(\theta,\theta'';t)}{f(\theta;t)},\\
l_*(\theta; \theta', \theta'', \theta''';t) &= \frac{l_2(\theta,\theta';t)\ l_2(\theta,\theta'';t)\ l_2(\theta,\theta''';t)}{f(\theta;t)^2}.
\end{align}
We call the resulting model the moment closure approximation (MCA) model.
The MCA model can be solved directly, and we can obtain a phase space, with ordered and disordered phases. In the case of a high rate of link dynamics, i.e. $d \to \infty$, the time evolution of the link density in \cref{eq:cont_l} leads to the steady state of 
\begin{equation}\label{eq:mean_field_closure}
    l_2(\theta,\theta';t)=\frac{c}{d}f(\theta;t)f(\theta';t).
\end{equation}
\Cref{eq:mean_field_closure} shows that for large sampling rates, the probability of two states being linked is proportional to the relative abundances of those states. In this case \cref{eq:cont_f} becomes independent of the link density and reduces to a single differential equation for $f(\theta;t)$.  We recognize this as the mean field approximation of the full model. 
We call the steady state solution of the mean field model $f_s(\theta)$, which is disordered if all states occur with same frequency i.e. $f_s(\theta)=\frac{1}{2 \pi}$. We call the solution ordered if one state is more abundant than all others, see \cref{fig:graphical_abstract}. To properly quantify this notion of order we introduce an order parameter $a_1$, which is the amplitude of the first symmetric Fourier mode in the series expansion of $f_s(\theta)$, given by
\begin{equation}\label{eq:f_fourier_exp}
    f_s(\theta)=\frac{1}{2\pi}\left[1+2\sum_{n\geq1} a_n \cos(n (\theta-\theta_s))\right].
\end{equation}
In \cref{eq:f_fourier_exp} the $a_n$ are the mode amplitudes and $\theta_s$ is the direction preferred in the steady state. Since $\cos(\theta-\theta_s)$ is a function with a single peak around $\theta_s$, the contribution of the first mode to the series expansion of $f_s(\theta)$ gives us a quantitative understanding of how ordered the system is. The atypical normalization of the Fourier series is chosen to have $0\leq a_1\leq 1$. Thus $a_1=0$ is associated with the disordered and $a_1=1$ with the ordered state.

\begin{figure}[b]
	\includegraphics[width=\linewidth]{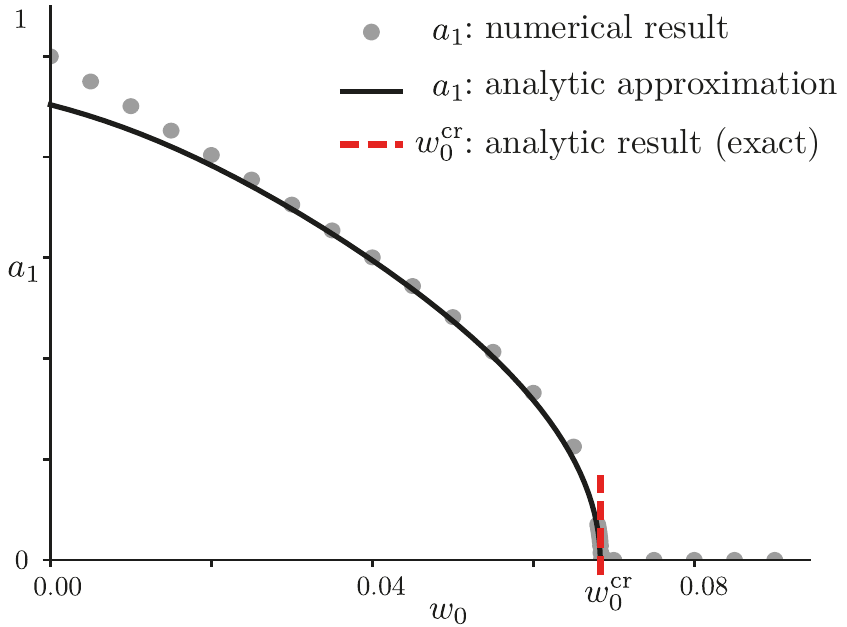}%
	\caption{\label{fig:MF_order_parameter} Bifurcation diagram of the system in the mean field approximation. The order parameter $a_1$ is plotted versus control parameter $w_0$. The dots represent numerical solutions, whereas the curve represents the best analytic approximation, after closing the system of Fourier coefficients at the fourth order.}
\end{figure}
\emph{\label{sec:results}Results -- }
We first solved the model in the mean field approximation, since in this case we only have an equation for $f(\theta;t)$ to solve, with $w_0$ as its only control parameter. Using the Fourier series expansion in \cref{eq:f_fourier_exp}, we can reduce the differential equation for the mean field system to a set of algebraic equations for the Fourier modes in steady state 
\begin{equation}\label{eq:amplitude_equation}
\partial_t a_n = \left( \frac{\sin(\frac{n\pi}{2})}{n \pi} -\frac{1}{4}-w_0\right)a_n +\sum_{p,q > 0}^{\infty}\frac{\gamma_{npq}}{(2\pi)^2}a_p a_q,
\end{equation}
where $\gamma_{npq}=$\\
$\int_{-\pi}^{\pi}\mathrm{d}\theta\int_{-\pi}^{\pi}\mathrm{d}\phi \cos\left[p\left(\theta+\frac{\phi}{2}\right)\right] \cos\left[q\left(\theta-\frac{\phi}{2}\right)\right]\cos\left(n\theta\right).$
From \cref{eq:amplitude_equation} we can extract the exact value of the critical noise 
\begin{equation}
    w^{\mathrm{cr}}_0 = \frac{1}{\pi} - \frac{1}{4},
\end{equation}
and the scaling behavior of the order parameter close to the critical point
$a_1 \propto (w_0-w^{\mathrm{cr}}_0)^{1/2}$. 
Our analytic approximation of $a_1$ is obtained by closing \cref{eq:amplitude_equation}. This procedure provides a generic $n$-th order polynomial for $a_1$ if closed at that order. Thus the best possible analytic solution is a root of a fourth order polynomial, which is plotted in \cref{fig:MF_order_parameter} as a solid black line. This solution gets progressively worse away from the critical point, but close to it our numerical and analytic results are in great agreement. Furthermore, we see that this phase transition is unambiguously second order and has a critical exponent of $1/2$, which was obtained analytically and is exact~\cite{supplement}.

\begin{figure}[b]
	\includegraphics[width=\linewidth]{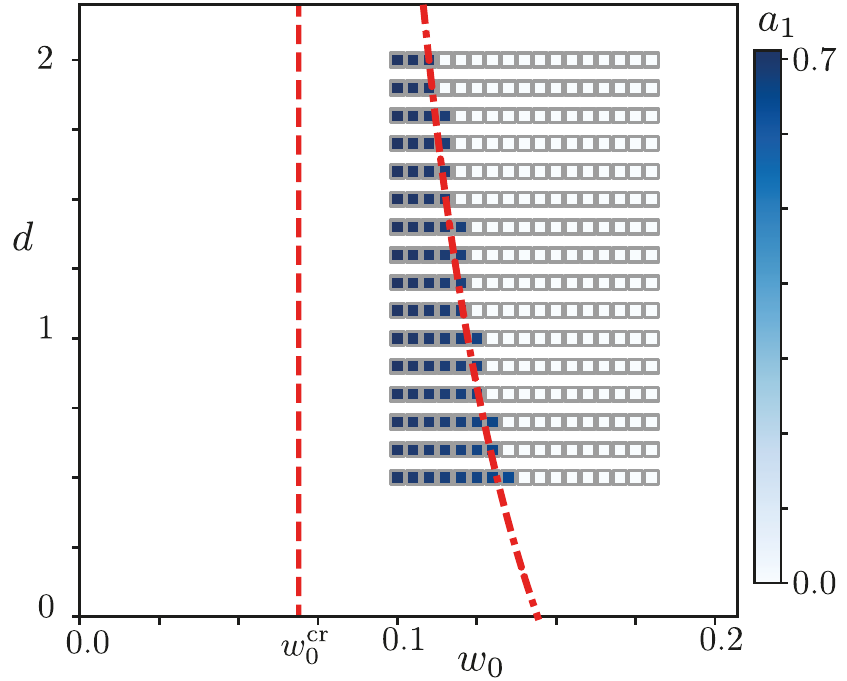}%
	\caption{Phase diagram for the MCA system in which both parameters $w_0$ and $d$ are varied. Squares represent numerical solutions. The curve is analytically determined and converges to $w_0^{\mathrm{cr}}$ for $d\to\infty$. Left from this curve the disordered state is linearly unstable which leads to $a_1>0$, such that the final system state is ordered. Right from this curve the disordered state is linearly stable, but the system can remain ordered (blue squares) if the initial condition of the system was ordered.}
	\label{fig:phase_diagram}
\end{figure}
For the MCA model we have an additional control parameter $d$, which sets the time scale of the link dynamics. Therefore, we get a critical line 
\begin{equation}
    d_{\mathrm{cr}}=-\frac{9}{4}-w_0+\frac{\pi -2} {2\pi(w_0-w_0^{\mathrm{cr}})}
\end{equation}
which can be seen in \cref{fig:phase_diagram}, as the red dash-dotted curve. The vertical red dashed line is drawn at the value $w_0^{\mathrm{cr}}$, this value is reached when $d$ goes to infinity. This critical line is obtained by linear stability analysis of the disordered state and is exact, just like the mean field model. The disordered state becomes unstable left of the critical line. Our numerical solutions were obtained by solving the system from an ordered initial condition. Thus the numerical solutions approach the phase transition from the ordered side. The existence of numerical results with nonzero order parameter, on the right of the line indicates hysteresis. Unfortunately, the methods that allowed us to obtain analytic values of the order parameter near criticality in the mean field model do not work here. However the numerical results in \cref{fig:phase_diagram}, clearly show a discontinuous change in the value of the order parameter at the critical line. Together with the bistable region, this is a clear sign of a first order phase transition. To exclude artifacts related to slow convergence to steady state, we repeated the runs for $3\times10^{5}$ and $9\times10^{5}$ time steps measured in units of $1/w_2$. The results remained unchanged~\cite{supplement}. 

\emph{\label{sec:Discussion}Discussion -- }
The continuous state adaptive network model can describe the emergence of swarming. The key parameter in this transition is the relative strength of the noise compared to the agent interactions. The critical noise value, at which the swarming transition occurs, can be changed by tuning the timescale of the network dynamics. However, if this timescale, given by $1/d$, becomes much smaller than the timescale of the node dynamics, set by $1/w_2$, two interesting things happen. First, the region of stability shrinks and the swarming phase transition changes from a first, to a second order transition. 
The shrinking of the stable region, for fast link dynamics, is surprising because, the limit of this regime is the mean field approximation. Using the intuition from equilibrium statistical mechanics, one would expect the critical noise to be higher in the case of the mean field model, since we expect mean field models to overestimate the tendency to order. However, we get the opposite result. The critical noise is maximal for $d=0$; most likely due to the decoupling of the link dynamics from the state dynamics. This decoupling can slow down the propagation of orientational information through the system.
The second striking effect seen in our system is the meta phase transition. While changing the rate of the link dynamics, the swarming transition becomes second order. Further investigation is necessary, but it seems that this effect can be explained by the observation that the type of noise affects the type of phase transition. According to Pimentel et al.~\cite{pimentel_intrinsic_2008} intrinsic noise in agent states leads to a second order transition, while extrinsic noise, caused by imperfect sampling of neighbors, leads to a first order phase transition. This fits well with our findings, since the mean field system only has intrinsic noise caused by spontaneous direction change, while the MCA system introduces an additional noise source, through changing topology. We do not know if this meta transition happens at a finite value of $d$ or strictly at $d\rightarrow\infty$, but regardless of the point where the meta transition happens, it constitutes a significant change in the behavior of the system. Biological systems that are described by this model, would therefore not always need nucleation points, such as small co-moving groups, for global swarms to emerge. If such systems could facilitate a quickly changing neighborhood topology, they would be able to smoothly transition into a swarm state. \\
The new model that we have put forward in this letter demonstrates that the timescale separation between the topology- and state-dynamics of the network has a strong impact on the nature of the swarming transition.
	\begin{acknowledgments}
	We thank Felix Frey for helpful discussions.
	G.D. was supported by the “BaSyC – Building a Synthetic Cell” Gravitation grant (024.003.019) of the Netherlands Ministry of Education, Culture and Science (OCW) and the Netherlands Organisation for Scientific Research (NWO).\\
	C.K. and G. D. contributed equally to this work.
	\end{acknowledgments}

%
	
\end{document}


\title{Supplemental material: Emergent order in a continuous state adaptive network model of living systems}
	\author{Carsten T. van de Kamp}
	\affiliation{Delft Institute of Applied Mathematics, Delft University of Technology, The Netherlands} 
	\affiliation{Department of Bionanoscience, Kavli Institute of Nanoscience, Delft University of Technology, The Netherlands}
	
	\author{George Dadunashvili}
	\affiliation{Department of Bionanoscience, Kavli Institute of Nanoscience, Delft University of Technology, The Netherlands}
	
	\author{Johan L.A. Dubbeldam}
	\email[]{J.L.A.Dubbeldam@TUDelft.nl}
	\affiliation{Delft Institute of Applied Mathematics, Delft University of Technology, The Netherlands}
	
	\author{Timon Idema}
	\email[]{T.Idema@TUDelft.nl}
	\affiliation{Department of Bionanoscience, Kavli Institute of Nanoscience, Delft University of Technology, The Netherlands}
	
	\date{\today}
	
	\maketitle
	\section{The governing equations for density evolution}
	Here we want to present the governing equations for the time evolution of the state density $f(\theta;t)$ and the link density $l_2(\theta,\theta';t)$, i.e. eq.~(1). These equations describe the dynamics according to the four rules, as given in the main text.
	Each rule has an associated type of dynamics and thus an associated term in the equations of both densities, which we indicate with a corresponding superscript. E.g. $\partial_t f^{(1)}(\theta;t)$ accounts for the effects of rule~1 on the time evolution of $f(\theta;t)$. Then the general form of our equations for state and link densities reads as follows:
	\begin{subequations}\label{seq:main_system_abstract_form}
		\begin{align}
			\partial_t f(\theta; t) &= \sum_{i=1}^{4}\partial_t f^{(i)}(\theta;t) \\
			\partial_t l_2(\theta, \theta'; t) &= \sum_{i=1}^{4}\partial_t l_2^{(i)}(\theta, \theta';t) 
		\end{align}
	\end{subequations}
	
	\subsection{Dynamics associated with rule~1}
	
	The effect of the dynamics of type~1 on the state density function $f(\theta; t)$ is captured by the following equation
	\begin{equation}
		\partial_t f^{(1)}(\theta;t)=w_0\int_{-\pi}^{\pi}\mathrm{d}\theta'\frac{1}{2\pi}\left[f(\theta';t) - f(\theta;t)\right]
	\end{equation}
	where $1/(2\pi)$ is the probability measure of a state change, which is a uniform distribution on the state space, since rule~1 prescribes random state changes that do not depend on the initial and final state. Using the fact that $f(\theta;t)$ is normalized we get
	\begin{equation}
		\partial_t f^{(1)}(\theta;t)=w_0\left(\frac{1}{2\pi} - f(\theta;t)\right).
	\end{equation}
	The change in link density due to interactions of type~1 is described by
	\begin{equation}
		\partial_t l_2^{(1)}(\theta,\theta';t) = w_0\int_{-\pi}^{\pi}\mathrm{d}\theta''\frac{1}{2\pi}\left[l_2(\theta,\theta'';t)+l_2(\theta',\theta'';t)-2 l_2(\theta,\theta';t)\right],
	\end{equation}
	where the first two terms account for the creation of a $\theta$ -- $\theta'$ pair, through the change of a single state. The last term accounts for the destruction of already existing $\theta$ -- $\theta'$ links, through the change of one of the states, which can happen two different ways, and hence the factor of two. Performing the integral on the last term gives
	\begin{equation}
		\partial_t l_2^{(1)}(\theta,\theta';t) = \frac{w_0}{2\pi}\int_{-\pi}^{\pi}\mathrm{d}\theta''\left[l_2(\theta,\theta'';t)+l_2(\theta',\theta'';t)\right]-2w_0 l_2(\theta,\theta';t).
	\end{equation}
	
	\subsection{Dynamics associated with rule~2}
	The next step will be to obtain similar equations for the second type of dynamics. For the state density $f(\theta;t)$, these interactions can be described as follows,
	\begin{align}
		\partial_t f^{(2)}(\theta;t)&=w_2\int_{-\pi}^{\pi}\mathrm{d}\theta'\int_{0}^{\frac{\pi}{2}}\mathrm{d}\xi\left[l_3(\theta';\theta-\xi,\theta+\xi;t) - l_3(\theta;\theta'-\xi,\theta'+\xi;t)\right].
	\end{align}
	First term represents nodes in an arbitrary state $\theta'$, in between two nodes that average to the state $\theta$. This way an extra $\theta$ node is created due to three body interactions at rate $w_2$. Second term corresponds to removal of $\theta$ nodes due to these interactions. We need to be careful that we take a proper circle average of the directions. This can be implemented in the equations by integrating over $\xi$ from 0 to $\pi/2$.
	The next differential equation captures the change in the link density function $\partial_t l_2(\theta,\theta';t)$ due to three body interactions
	\begin{subequations}
		\begin{align}
			\partial_t l_2^{(2)}(\theta,\theta';t) =&\nonumber \\ 
			&w_2 \int_{-\pi}^{\pi}\mathrm{d}\theta''\int_{0}^{\pi/2}\mathrm{d}\xi\; l_*(\theta'';\theta,\theta'+\xi,\theta'-\xi;t) \label{seq:l22_1} \\ 
			&+ w_2 \int_{-\pi}^{\pi}\mathrm{d}\theta''\int_{0}^{\pi/2}\mathrm{d}\xi\; l_*(\theta'';\theta',\theta+\xi,\theta-\xi;t) \label{seq:l22_2} \\
			&- w_2 \int_{-\pi}^{\pi}\mathrm{d}\theta''\int_{0}^{\pi/2}\mathrm{d}\xi\; l_*(\theta;\theta',\theta''+\xi,\theta''-\xi;t) \label{seq:l22_3} \\
			&- w_2 \int_{-\pi}^{\pi}\mathrm{d}\theta''\int_{0}^{\pi/2}\mathrm{d}\xi\; l_*(\theta';\theta,\theta''+\xi,\theta''-\xi;t) \label{seq:l22_4}\\
			&+w_2 \int_{-\pi}^{\pi}\mathrm{d}\theta''\; l_3(\theta'';\theta,-\theta+2\theta';t) \label{seq:l22_5}\\
			&+w_2 \int_{-\pi}^{\pi}\mathrm{d}\theta''\; l_3(\theta'';\theta',-\theta'+2\theta;t) \label{seq:l22_6} \\
			&-w_2 \int_{-\pi}^{\pi}\mathrm{d}\theta''\; l_3(\theta;\theta',\theta'';t) \label{seq:l22_7}\\
			&-w_2 \int_{-\pi}^{\pi}\mathrm{d}\theta''\; l_3(\theta';\theta,\theta'';t) \label{seq:l22_8}.
		\end{align}
	\end{subequations}
	The \cref{seq:l22_1} and \cref{seq:l22_2} account for four-point subgraphs where one of the nodes is already pointing in the right direction $\theta$ or $\theta'$ and the center one can average with the remaining two neighbors to change into $\theta'$ or $\theta$ respectively and thus create a $\theta$ -- $\theta'$ link. The terms of \cref{seq:l22_3} and \cref{seq:l22_4}, describe situations where a $\theta$ -- $\theta'$ link already exists in a star like subgraph and gets destroyed. Again, we integrate over all $\xi$ for which our averaging operation is defined properly. Three-point subgraphs with the potential to form an $\theta$--$\theta'$ link are taken into account in \cref{seq:l22_5} and \cref{seq:l22_6}. In these cases $\theta''$ changes to $\theta$ or $\theta'$, while the one of the neighbor nodes is already in $\theta'$ or $\theta$ configuration respectively. Finally, \cref{seq:l22_7} and \cref{seq:l22_8} account the loss term on the three-point subgraphs level.
	\subsection{Dynamics associated with rule~3 and rule~4}
	Since the dynamics described in rule~3 and rule~4 only affect links, 
	\begin{equation}
		\partial_t f^{(3)}(\theta;t) =  \partial_t f^{(4)}(\theta;t) = 0.
	\end{equation}
	Link creation dynamics, between two nodes of type $\theta$ and $\theta'$ is given by the density of already existing nodes and the link creation rate $c$
	\begin{equation}
		\partial_t l_2^{(3)}(\theta,\theta';t)=c f(\theta;t)f(\theta';t).
	\end{equation}
	Link deletion dynamics is given by the number of already existing links and the deletion rate $d$
	\begin{equation}
		\partial_t l_2^{(4)}(\theta,\theta';t)=-d l_2(\theta,\theta';t).
	\end{equation}
	
	\subsection{Full equations}
	Putting everything together in \cref{seq:main_system_abstract_form} we arrive at 
	\begin{subequations}\label{seq:f_full_form}
		\begin{align}
			\partial_t f(\theta; t) &= w_0\left(\frac{1}{2\pi} - f(\theta;t)\right) + w_2\int_{-\pi}^{\pi}\mathrm{d}\theta'\int_{0}^{\frac{\pi}{2}}\mathrm{d}\xi\left[l_3(\theta';\theta-\xi,\theta+\xi;t) - l_3(\theta;\theta'-\xi,\theta'+\xi;t)\right],
		\end{align}
	\end{subequations}
	for the state density distribution and we can identify the definition of the functional $F^{\mathrm{int}}[l_3;\theta;t]$ from the second term as
	\begin{equation}
		F^{\mathrm{int}}[l_3;\theta;t]=\int_{-\pi}^{\pi}\mathrm{d}\theta'\int_{0}^{\frac{\pi}{2}}\mathrm{d}\xi\left[l_3(\theta';\theta-\xi,\theta+\xi;t) - l_3(\theta;\theta'-\xi,\theta'+\xi;t)\right].
	\end{equation}
	For the link density distribution we get
	
	\begin{subequations}\label{seq:l_full_form}
		\begin{flalign}
			\partial_t l_2^{(2)}(\theta,\theta';t) =&\nonumber \\ 
			& \frac{w_0}{2\pi}\int_{-\pi}^{\pi}\mathrm{d}\theta'' \big[l_2(\theta,\theta'';t)+l_2(\theta',\theta'';t)  \big]-2w_0 l_2(\theta,\theta';t)  \label{seq:L_noise}\\
			&+ w_2\int_{-\pi}^{\pi}\mathrm{d}\theta'' \Bigg\{
			l_3(\theta'';\theta,-\theta+2\theta';t) + l_3(\theta'';\theta',-\theta'+2\theta;t) - l_3(\theta;\theta',\theta'';t) - l_3(\theta';\theta,\theta'';t) & \nonumber\\
			& \ \ \ \ \ \ \ \ \ \ \ \ \ \ \ \ \ \ \ \ \ \ \ \ + \int_{0}^{\pi/2}\mathrm{d}\xi 
			\big[l_*(\theta'';\theta,\theta'+\xi,\theta'-\xi;t) + l_*(\theta'';\theta',\theta+\xi,\theta-\xi;t)  \nonumber \\
			& \ \ \ \ \ \ \ \ \ \ \ \ \ \ \ \ \ \ \ \ \ \ \ \ \ \ \ \ \ \ \ \ \ \ \ \ \ \ \ \ \ \  - l_*(\theta;\theta',\theta''+\xi,\theta''-\xi;t) - l_*(\theta';\theta,\theta''+\xi,\theta''-\xi;t)\big] \Bigg\} \label{seq:L_int} \\
			&+cf(\theta;t)f(\theta';t)-dl_2(\theta,\theta';t).
		\end{flalign}
	\end{subequations}
	The functional $w_0 L^{\mathrm{noise}}$ is given by \cref{seq:L_noise} and $w_2 L^{\mathrm{int}}$ is given by \cref{seq:L_int}. As mentioned in the main text $L^{\mathrm{noise}}$ and $L^{\mathrm{int}}$ integrate to zero. We demonstrate this later in \cref{ssec:properties_of_L}, but it should already be conceptually clear, since these two functionals account for conservative dynamics that come from the state change. No links are created or destroyed during this type of changes, and thus we do not expect this to cause any global changes in the overall link density.
	
	\subsection{Properties of $L^{\mathrm{noise}}$ and $L^{\mathrm{int}}$}\label{ssec:properties_of_L}
	We first show that $L^{\mathrm{noise}}$ integrates to zero. This is done by straight forward integration,
	\begin{align*}
		\int_{-\pi}^{\pi}\mathrm{d}\theta\int_{-\pi}^{\pi}\mathrm{d}\theta'L^{\mathrm{noise}}[l_2;\theta,\theta';t]&= \\
		&\frac{1}{2\pi}\int_{-\pi}^{\pi}\mathrm{d}\theta\int_{-\pi}^{\pi}\mathrm{d}\theta'\int_{-\pi}^{\pi}\mathrm{d}\theta'' \big[l_2(\theta,\theta'';t)+l_2(\theta',\theta'';t)  \big]- 2 \int_{-\pi}^{\pi}\mathrm{d}\theta\int_{-\pi}^{\pi}\mathrm{d}\theta'l_2(\theta,\theta';t) \\
		&=k(t)+k(t)-2k(t) =0,
	\end{align*}
	where we used the definition of $k(t)$, as provided in eq. (2) of the main text. The proof that $L^{\mathrm{int}}$ integrates to zero is also a straight forward integration. Writing everything out
	\begin{subequations}
		\begin{align}
			\int_{-\pi}^{\pi}\mathrm{d}\theta\int_{-\pi}^{\pi}\mathrm{d}\theta'L^{\mathrm{int}}[l_3,l_*;\theta,\theta';t]=& \nonumber \\ 
			&\int_{-\pi}^{\pi}\mathrm{d}\theta\int_{-\pi}^{\pi}\mathrm{d}\theta' \int_{-\pi}^{\pi}\mathrm{d}\theta''\int_{0}^{\pi/2}\mathrm{d}\xi l_*(\theta'';\theta,\theta'+\xi,\theta'-\xi;t) \label{seq:L_int_22_1} \\ 
			&+ \int_{-\pi}^{\pi}\mathrm{d}\theta\int_{-\pi}^{\pi}\mathrm{d}\theta' \int_{-\pi}^{\pi}\mathrm{d}\theta''\int_{0}^{\pi/2}\mathrm{d}\xi l_*(\theta'';\theta',\theta+\xi,\theta-\xi;t) \label{seq:L_int_22_2} \\
			&- \int_{-\pi}^{\pi}\mathrm{d}\theta\int_{-\pi}^{\pi}\mathrm{d}\theta' \int_{-\pi}^{\pi}\mathrm{d}\theta''\int_{0}^{\pi/2}\mathrm{d}\xi l_*(\theta;\theta',\theta''+\xi,\theta''-\xi;t) \label{seq:L_int_22_3} \\
			&- \int_{-\pi}^{\pi}\mathrm{d}\theta\int_{-\pi}^{\pi}\mathrm{d}\theta' \int_{-\pi}^{\pi}\mathrm{d}\theta''\int_{0}^{\pi/2}\mathrm{d}\xi l_*(\theta';\theta,\theta''+\xi,\theta''-\xi;t) \label{seq:L_int_22_4}\\
			&+\int_{-\pi}^{\pi}\mathrm{d}\theta\int_{-\pi}^{\pi}\mathrm{d}\theta' \int_{-\pi}^{\pi}\mathrm{d}\theta'' l_3(\theta'';\theta,-\theta+2\theta';t) \label{seq:L_int_22_5}\\
			&+\int_{-\pi}^{\pi}\mathrm{d}\theta\int_{-\pi}^{\pi}\mathrm{d}\theta' \int_{-\pi}^{\pi}\mathrm{d}\theta'' l_3(\theta'';\theta',-\theta'+2\theta;t) \label{seq:L_int_22_6} \\
			&-\int_{-\pi}^{\pi}\mathrm{d}\theta\int_{-\pi}^{\pi}\mathrm{d}\theta' \int_{-\pi}^{\pi}\mathrm{d}\theta'' l_3(\theta;\theta',\theta'';t) \label{seq:L_int_22_7}\\
			&-\int_{-\pi}^{\pi}\mathrm{d}\theta\int_{-\pi}^{\pi}\mathrm{d}\theta' \int_{-\pi}^{\pi}\mathrm{d}\theta'' l_3(\theta';\theta,\theta'';t) \label{seq:L_int_22_8},
		\end{align}
	\end{subequations}
	one can see that one gets pairs of equations \cref{seq:L_int_22_1}, \cref{seq:L_int_22_3}; \cref{seq:L_int_22_2}, \cref{seq:L_int_22_4}; \cref{seq:L_int_22_5}, \cref{seq:L_int_22_7} and \cref{seq:L_int_22_6}, \cref{seq:L_int_22_8} that only differ by relabeling of their integration variables but have opposite signs, so everything adds up to zero.
	
	\section{Moment closure approximation}
	In this section we derive eq.~(3) and eq.~(4) from the main text. Moment closure effectively means that the probability of finding higher order structures decomposes into the product of the probabilities of finding lower order structures. For triplets of $\theta'$ -- $\theta$ -- $\theta''$ where $\theta$ is the middle node, like depicted in figure (2f) of the main text, this means
	\begin{subequations}
		\begin{align}
			p(\theta;\theta',\theta'';t)&=p(\theta',\theta''|\theta;t)p(\theta;t) \label{seq:l3_product_rule}\\
			&=p(\theta'|\theta'',\theta;t)p(\theta''|\theta;t)p(\theta;t)\label{seq:l3_decoupling_probabilities} \\
			&=p(\theta'|\theta;t)p(\theta''|\theta;t)p(\theta;t)\label{seq:l3_decoupled_probabilities},
		\end{align}
	\end{subequations}
	where $p(\theta;\theta',\theta'';t)$ is the full probability of finding the triplet $\theta'$ -- $\theta$ -- $\theta''$, at time $t$. Note that the semicolon after the first argument does not denote a conditional probability, but it simply separates the central node from the peripheral nodes. $p(\theta',\theta''|\theta)$ is the conditional probability that the peripheral nodes have the angles $\theta'$ and $\theta''$, given that the central node has the angle $\theta$. $p(\theta;t)$ is the prior probability of finding a node, which has the angle $\theta$, i.e. $p(\theta;t)=f(\theta;t)$. Going from \cref{seq:l3_decoupling_probabilities} to \cref{seq:l3_decoupled_probabilities} we made the closure assumption $p(\theta'|\theta'',\theta;t)=p(\theta'|\theta;t)$, saying that the probability of finding one peripheral node does not depend on the state of the other peripheral node. Now we simply relate the probability of finding a state, to the density of states
	\begin{align}
		\frac{l_3(\theta;\theta',\theta'';t)}{\|l_3\|}&=p(\theta;\theta',\theta'';t)\\
		\frac{l_2(\theta;\theta';t)}{k}&=p(\theta;\theta';t)=p(\theta'|\theta;t)p(\theta;t),\\
	\end{align}
	where $k$ is the norm of the link density as defined in eq.~(2) of the main text and $||l_3||$ is the norm of the triplet density, defined as
	\begin{align}
		\|l_3\|=\int_{-\pi}^{\pi}\mathrm{d}\theta\int_{-\pi}^{\pi}\mathrm{d}\theta' \int_{-\pi}^{\pi}\mathrm{d}\theta''l_3(\theta;\theta',\theta'';t).
	\end{align}
	Putting everything together, we get
	\begin{align}
		l_3(\theta;\theta',\theta'';t)&=\frac{\|l_3\|}{k^2}\frac{l_2(\theta,\theta';t)l_2(\theta,\theta'';t)}{f(\theta;t)}.\label{seq:l_3_closure}
	\end{align}
	Repeating the same steps for the star like, four point subgraphs we get,
	\begin{align}\label{seq:l_star_closure_derivation}
		\frac{l_*(\theta;\theta',\theta'',\theta''';t)}{\|l_*\|}&=p(\theta;\theta',\theta'',\theta''';t) \nonumber \\
		&=p(\theta',\theta'',\theta'''|\theta;t)p(\theta;t) \nonumber \\
		&=p(\theta'|\theta;t)p(\theta''|\theta;t)p(\theta'''|\theta;t)p(\theta;t) \nonumber \\
		&=\frac{p(\theta',\theta;t)}{p(\theta;t)}\frac{p(\theta'',\theta;t)}{p(\theta;t)}\frac{p(\theta''',\theta;t)}{p(\theta;t)}p(\theta;t) \nonumber \\
		&=\frac{l_2(\theta,\theta';t)l_2(\theta,\theta'';t)l_2(\theta,\theta''';t)}{f^2(\theta;t)k^3},
	\end{align}
	where $\|l_*\|$ is the norm of the density of star like subgraphs and is defined as
	\begin{equation}
		\|l_*\|=\int_{-\pi}^{\pi}\mathrm{d}\theta\int_{-\pi}^{\pi}\mathrm{d}\theta' \int_{-\pi}^{\pi}\mathrm{d}\theta''\int_{-\pi}^{\pi}\mathrm{d}\theta''' \;l_*(\theta;\theta',\theta'',\theta''';t).
	\end{equation}
	Rearranging terms in \ref{seq:l_star_closure_derivation} gets us to,
	\begin{align}\label{seq:l_star_closure}
		l_*(\theta;\theta',\theta'',\theta''';t)&=
		\frac{\|l_*\|}{k^3}\frac{l_2(\theta,\theta';t)l_2(\theta,\theta'';t)l_2(\theta,\theta''';t)}{f^2(\theta;t)}.
	\end{align}
	The only differences between the here derived \cref{seq:l_3_closure} and \cref{seq:l_star_closure} and the closure relations given in eq.~(4) and eq.~(5) of the main text, are the prefactors of $\|l_3\|/k^2$ and $\|l_*\|/k^3$. These prefactors represent the topology of the underlying graphs, on which our dynamics take place. $k$ is the link density, i.e. the number of links in the network relative to the number of nodes, and since each link is shared by two nodes, we can estimate the average node degree $n=2k$. $\|l_3\|$ is the number of triplets a node is the center of and $\|l_*\|$ is the number of quadruplets a node is the center of. In a fully connected graph with the node degree $n$ each node will be the center of $n\choose2$ triplets and $n\choose3$ quadruplets. This means that the upper bounds on our prefactors are,
	\begin{align}
		\frac{\|l_3\|}{k^2}&\leq\frac{n(n-1)/2}{(n/2)^2}=2+\mathcal{O}(n^{-1}) \\
		\frac{\|l_*\|}{k^3}&\leq\frac{n(n-1)(n-2)/6}{(n/2)^3}=\frac{4}{3}+\mathcal{O}(n^{-1}).
	\end{align}
	The actual values will be dependent on the exact realization of the random network topology, and one needs to make an assumption at this point. For our model we set both of these ratios to one,
	\begin{align}
		\frac{\|l_3\|}{k^2}&=1 \\
		\frac{\|l_*\|}{k^3}&=1.
	\end{align}
	Since these values are below the fully connected network bounds, we interpret them as equivalent to choosing a sparse network topology.
	
	Using the moment closure relations as given in eq.~(4) and eq.~(5) of the main text, we get a closed system of differential equations for the state and link densities,
	\begin{subequations}\label{seq:mca_system}
		\begin{flalign}
			\partial_t f(\theta; t) =& w_0\left(\frac{1}{2\pi} - f(\theta;t)\right) \nonumber \\
			&+ w_2\int_{-\pi}^{\pi}\mathrm{d}\theta'\int_{0}^{\frac{\pi}{2}}\mathrm{d}\xi \left[\frac{l_2(\theta',\theta-\xi;t)l_2(\theta',\theta+\xi;t)}{f(\theta';t)} - \frac{l_2(\theta,\theta'-\xi;t)l_2(\theta,\theta'+\xi;t)}{f(\theta;t)}\right] \label{seq:f_mca} \\
			\partial_t l_2^{(2)}(\theta,\theta';t) =&\nonumber \\ 
			& \frac{w_0}{2\pi}\int_{-\pi}^{\pi}\mathrm{d}\theta'' \big[l_2(\theta,\theta'';t)+l_2(\theta',\theta'';t)  \big]-2w_0 l_2(\theta,\theta';t)  \nonumber\\
			&+ w_2\int_{-\pi}^{\pi}\mathrm{d}\theta'' \Bigg\{
			\frac{l_2(\theta'',\theta;t)l_2(\theta'',-\theta+2\theta';t)}{f(\theta'';t)} +
			\frac{l_2(\theta'',\theta';t)l_2(\theta'',-\theta'+2\theta;t)}{f(\theta'';t)} \nonumber \\
			& \ \ \ \ \ \ \ \ \ \ \ \ \ \ \ \ \ \ \ \ \ \ \ \ - \frac{l_2(\theta,\theta';t)l_2(\theta,\theta'';t)}{f(\theta;t)}
			- \frac{l_2(\theta',\theta;t)l_2(\theta',\theta'';t)}{f(\theta';t)}& \nonumber \nonumber \\
			& \ \ \ \ \ \ \ \ \ \ \ \ \ \ \ \ \ \ \ \ \ \ \ \ + \int_{0}^{\pi/2}\mathrm{d}\xi
			\big[
			\frac{l_2(\theta'',\theta;t)l_2(\theta'',\theta'+\xi;t)l_2(\theta'',\theta'-\xi;t)}{f^2(\theta'';t)} \nonumber \\
			& \ \ \ \ \ \ \ \ \ \ \ \ \ \ \ \ \ \ \ \ \ \ \ \ \ \ \ \ \ \ \ \ \ \ \ \ \ \ \ \ \ \ \ \ \ 
			+ \frac{l_2(\theta'',\theta';t)l_2(\theta'',\theta+\xi;t)l_2(\theta'',\theta-\xi;t)}{f^2(\theta'';t)} \nonumber \\
			& \ \ \ \ \ \ \ \ \ \ \ \ \ \ \ \ \ \ \ \ \ \ \ \ \ \ \ \ \ \ \ \ \ \ \ \ \ \ \ \ \ \ \ \ \ 
			-\frac{l_2(\theta,\theta';t)l_2(\theta,\theta''+\xi;t)l_2(\theta,\theta''-\xi;t)}{f^2(\theta;t)} \nonumber \\
			& \ \ \ \ \ \ \ \ \ \ \ \ \ \ \ \ \ \ \ \ \ \ \ \ \ \ \ \ \ \ \ \ \ \ \ \ \ \ \ \ \ \ \ \ \ -\frac{l_2(\theta',\theta;t)l_2(\theta',\theta''+\xi;t)l_2(\theta',\theta''-\xi;t)}{f^2(\theta';t)}\big]\Bigg\} \nonumber \\
			&+cf(\theta;t)f(\theta';t)- d l_2(\theta,\theta';t).\label{seq:l_mca}
		\end{flalign}
	\end{subequations}
	We used \cref{seq:mca_system} to obtain the numeric results for the MCA model, since it has good numerical stability. But it is possible to simplify the equations further by exploiting the symmetry of $l_2$ in its arguments and factor many terms.
	Which we will make use of in \cref{ssec:mf_derivations} and \cref{ssec:mca_derivations} to find the phase transition analytically.
	
	\section{Mean field approximation}
	Since the closed system from \cref{seq:mca_system} is still difficult to analyze, both numerically and analytically, one can simplify the model further and close it on the level of the state density $f(\theta;t)$. This is a good approximation for diverging rates of link dynamics i.e. $d\to\infty$ and $c\to\infty$, while $c/d=k_s=\mathrm{const}$. Then the last two terms in \cref{seq:l_mca} dominate the dynamics of $l_2$ and lead to the steady state,
	\begin{equation}
		l_2(\theta,\theta';t)=\frac{c}{d}f(\theta;t)f(\theta';t)=k_sf(\theta;t)f(\theta';t),\label{seq:mean_field_approx}
	\end{equation}
	where $l_2$ is directly determined by the state density $f$, which can still change through it's own dynamics. Substituting \cref{seq:mean_field_approx} into \cref{seq:f_mca} we get,
	\begin{equation}
		\partial_t f(\theta;t) = w_0 \left( \frac{1}{2\pi} -f(\theta;t)\right) + w_2k_s^2\left\{\int_{0}^{\frac{\pi}{2}}\mathrm{d}\xi \; f(\theta-\xi;t)f(\theta+\xi;t) - f(\theta;t)\int_{-\pi}^{\pi}\mathrm{d}\theta'\int_{0}^{\frac{\pi}{2}}\mathrm{d}\xi \; f(\theta'-\xi;t)f(\theta'+\xi;t) \right\}. \label{seq:mf_long}
	\end{equation}
	We used \cref{seq:mf_long} for numerical solutions, because it is numerically stable. But if we recognize that 
	\begin{equation}\label{seq:integralis_egregium}
		\int_{-\pi}^{\pi}\mathrm{d}\theta'\int_{0}^{\frac{\pi}{2}}\mathrm{d}\xi\; f(\theta'-\xi;t)f(\theta'+\xi;t) =\frac{1}{4},
	\end{equation}
	then, the mean field equation simplifies to
	\begin{equation}\label{seq:mf_short}
		\partial_t f(\theta;t)= \frac{w_0}{2\pi} - \left(w_0 + \frac{w_2k_s^2}{4}\right) f(\theta;t) +\frac{w_2k_s^2}{4}\int_{-\pi}^{\pi}\mathrm{d}\xi\; f\left(\theta-\frac{\xi}{2};t\right)f\left(\theta+\frac{\xi}{2};t\right),
	\end{equation}
	Where we used \cref{seq:xi_symmetry} to change the integration boundary of the $\xi$ integral and \cref{seq:integralis_egregium} is a special case of \cref{seq:integralis_egregium_generalis}.
	\section{Solution of the mean field model}\label{ssec:mf_derivations}
	\subsection{Critical noise}
	To obtain the value of the critical noise, $w_0^{\mathrm{cr}}$ we perform the linear stability analysis of \cref{seq:mf_short}, by expanding the state density around the uniform state, in Fourier modes. In our simulations we had observed that the steady state distribution is always symmetric around the mean, thus we used the expansion in symmetric modes, as shown in eq.~(7) of the main text. Here we also set $\theta_s=0$, without loss of generality, and start from
	\begin{equation}\label{seq:fourier_expansion}
		f(\theta;t)=\frac{1}{2\pi}\left[1+2\sum_{n=1}^{\infty}a_n \cos(n\theta)\right].
	\end{equation}
	Plugging \cref{seq:fourier_expansion} in \cref{seq:mf_short} we get
	
	\begin{align*}
		2\sum_{n=1}^{\infty} \partial_t a_n \cos(n\theta) =& -\frac{w_2k_s^2}{4}- \left(w_0+\frac{w_2k_s^2}{4} \right) 2\sum_{n=1}^{\infty}a_n \cos(n\theta) \\
		&+ \frac{w_2k_s^2}{4}\frac{1}{2\pi}\int_{-\pi}^{\pi}\mathrm{d}\xi\left[1+2\sum_{n=1}^{\infty}a_n \cos\left(n\left(\theta+\frac{\xi}{2}\right)\right)\right]\left[1+2\sum_{m=1}^{\infty}a_m \cos\left(m\left(\theta-\frac{\xi}{2}\right)\right)\right] \nonumber\\
		=&-\left(w_0 + \frac{w_2k_s^2}{4} \right) 2\sum_{n=1}^{\infty}a_n \cos(n\theta) + 
		\frac{w_2k_s^2}{4}\frac{1}{2\pi}2\sum_{n=1}^{\infty}a_n\;2\int_{-\pi}^{\pi}\mathrm{d}\xi\cos\left(n\left(\theta+\frac{\xi}{2}\right)\right) \nonumber\\
		&+ \frac{w_2k_s^2}{4}\frac{4}{2\pi}\sum_{m,n=1}^{\infty}a_n a_m \int_{-\pi}^{\pi}\mathrm{d}\xi \cos\left(n\left(\theta+\frac{\xi}{2}\right)\right)\cos\left(m\left(\theta-\frac{\xi}{2}\right)\right),
	\end{align*}
	canceling the factor of two on both sides and using,
	\begin{equation}
		\int_{-\pi}^{\pi}\mathrm{d}\xi\cos\left(n\left(\theta+\frac{\xi}{2}\right)\right)=\frac{4\sin\left(\frac{n\pi}{2}\right)}{n}\cos(n\theta)\nonumber,
	\end{equation}
	we get,
	\begin{align*}
		\sum_{n=1}^{\infty} \partial_t a_n \cos(n\theta) =&
		-\left(w_0 + \frac{w_2k_s^2}{4} \right) \sum_{n=1}^{\infty}a_n \cos(n\theta) + 
		\frac{w_2k_s^2}{4}\sum_{n=1}^{\infty}a_n\;\frac{4\sin\left(\frac{n\pi}{2}\right)}{n\pi}\cos(n\theta)\\
		&+ \frac{w_2k_s^2}{4\pi}\sum_{m,n=1}^{\infty}a_n a_m \int_{-\pi}^{\pi}\mathrm{d}\xi \cos\left(n\left(\theta+\frac{\xi}{2}\right)\right)\cos\left(m\left(\theta-\frac{\xi}{2}\right)\right).
	\end{align*}
	Finally to get rid of the sums we multiply both sides with $\cos(p\theta)$ and integrate over $\theta$. This allows us to use
	\begin{equation}
		\int_{-\pi}^{\pi}\mathrm{d}\theta \cos(n\theta)\cos(p\theta) = \pi \delta_{np},
	\end{equation}
	to get the nonlinear amplitude equation
	\begin{align}\label{seq:amplitude_equation}
		\partial_t a_n =&
		\left(w_2k_s^2\left(\frac{\sin\left(\frac{n\pi}{2}\right)}{n\pi}-\frac{1}{4}\right)-w_0 \right) a_n + \frac{w_2k_s^2}{(2\pi)^2}\sum_{p,q=1}^{\infty}a_p a_q \gamma_{npq},
	\end{align}
	where we relabeled the indices $p$, $n$, $m$ to $n$, $p$, $q$, and introduced
	\begin{equation*}
		\gamma_{npq}=\int_{-\pi}^{\pi}\mathrm{d}\theta\int_{-\pi}^{\pi}\mathrm{d}\xi \cos\left(p\left(\theta+\frac{\xi}{2}\right)\right)\cos\left(q\left(\theta-\frac{\xi}{2}\right)\right).
	\end{equation*}
	To determine the linear stability of the disordered phase we only need to look at the linear part of \cref{seq:amplitude_equation},
	\begin{align}\label{seq:amplitude_equation_linearized}
		\partial_t a_n = (w_0^{\mathrm{cr}}(n)-w_0) a_n.
	\end{align}
	With $w_0^{\mathrm{cr}}(n)=w_2k_s^2\left(\frac{\sin\left(\frac{n\pi}{2}\right)}{n\pi}-\frac{1}{4}\right)$. The solution of \cref{seq:amplitude_equation_linearized} is 
	\begin{equation}
		a_n(t)=a_n(0)\exp\left[(w_0^{\mathrm{cr}}(n)-w_0)t\right].
	\end{equation}\label{seq:amplitude_equation_linearized_solution}
	If $(w_0^{\mathrm{cr}}(n)-w_0)>0$ for any integer $n\geq1$ than the amplitude of that mode will start growing and will destabilize the disordered state. For this to happen $w_0^{\mathrm{cr}}(n)$ has to be positive in the first place, which is only the case for $n=1$. Then the condition for the critical noise becomes
	\begin{equation}
		w_2k_s^2\left(\frac{1}{\pi}-\frac{1}{4}\right)-w_0>0,
	\end{equation}
	and we get the general result for the critical noise
	\begin{equation}
		w_0^{\mathrm{cr}}=w_2k_s^2\left(\frac{1}{\pi}-\frac{1}{4}\right).
	\end{equation}
	\subsection{Approximate solution of the nonlinear equation}
	We start by evaluating integral for $\gamma_{npq}$, which leads to
	\begin{equation}
		\gamma_{npq}=\frac{2\pi}{p^2-q^2}\left(\zeta_{npq}-\zeta_{nqp}\right),
	\end{equation}
	where
	\begin{subequations}
		\begin{align}
			\zeta_{npq} &= \cos\left( \frac{p\pi}{2}\right) \sin\left( \frac{q\pi}{2}\right)\bigg[ (p-q)\left( \delta_{p,q-n} +\delta_{p,q+n}\right)-(p+q)\left( \delta_{p,-q+n} + \delta_{p,-q-n}\right)\bigg]\\
			&=\frac{1}{2}\bigg[\sin\left((p+q)\frac{\pi}{2}\right)-\sin\left((p-q)\frac{\pi}{2}\right)\bigg]\bigg[ (p-q)\left( \delta_{p,q-n} +\delta_{p,q+n}\right) - (p+q) \delta_{p,-q+n}\bigg].\label{seq:zeta2}
		\end{align}
	\end{subequations}
	In \cref{seq:zeta2} we dropped $\delta_{p,-q-n}$ since this would lead to $a_{-p-n}$ which are zero because all indices must be positive. Plugging the definitions of $\gamma_{npq}$ and $\zeta_{npq}$ in the nonlinear part of \cref{seq:amplitude_equation} we get
	\begin{subequations}
		\begin{align}
			\frac{1}{(2\pi)^2}\sum_{p,q} a_p a_q\gamma_{npq}&=
			\frac{1}{(2\pi)^2}\sum_{p,q} a_p a_q \frac{2\pi}{p^2-q^2}\left(\zeta_{npq}-\zeta_{nqp}\right).\label{seq:ampl_eqn_nonlinear_part_simplification_1}\\
			&=\frac{1}{2\pi}\sum_{p,q=1}^{\infty} a_p a_q \frac{2\zeta_{npq}}{p^2-q^2}\\
			&=\frac{1}{2\pi}\sum_{p,q=1}^{\infty} a_p a_q \bigg[\sin\left((p+q)\frac{\pi}{2}\right)-\sin\left((p-q)\frac{\pi}{2}\right)\bigg] \left[ \frac{\delta_{p,q-n} +\delta_{p,q+n}}{p+q}-\frac{\delta_{p,-q+n}}{p-q}\right]\\
			&=\frac{\sin\left(\frac{n\pi}{2}\right)}{2\pi}\sum_{q=1}^{\infty} a_q\left(\frac{(-1 + (-1)^q)a_{n-q}}{n-2q}+\frac{\left(1-(-1)^q\right)a_{q-n}}{2q-n}+\frac{(-1 + (-1)^q)a_{q+n}}{n+2q}\right)\label{seq:ampl_eqn_nonlinear_part_simplification_2}\\
			&=\frac{\sin\left(\frac{n\pi}{2}\right)}{2\pi} \left(\sum_{q=1}^{n-1}\frac{(-1 + (-1)^q)a_{n-q}a_q}{n-2q}+\sum_{q=n+1}^{\infty}\frac{\left(1-(-1)^q\right)a_{q-n}a_q}{2q-n}+\sum_{q=1}^{\infty}\frac{(-1 + (-1)^q)a_{q+n}a_q}{n+2q}\right)\label{seq:ampl_eqn_nonlinear_part_simplification_3} \\
			&=\frac{\sin\left(\frac{n\pi}{2}\right)}{2\pi} \left(\sum_{q=1}^{n-1}\frac{(-1 + (-1)^q)a_{n-q}a_q}{n-2q}+(1 - (-1)^n)\sum_{q=1}^{\infty}\frac{(-1)^q a_{n+q}a_q}{n+2q}\right)\label{seq:ampl_eqn_nonlinear_part_simplification_4}.
		\end{align}
		We relabeled $p$ and $q$ in the second term of \cref{seq:ampl_eqn_nonlinear_part_simplification_1}. In \cref{seq:ampl_eqn_nonlinear_part_simplification_3} we again make use of the property $a_{n}=0$ $\forall n\leq0$ to constrain the boundaries of the first two sums. After shifting the index in the second sum $p\to p+n$ we arrive at the final form \cref{seq:ampl_eqn_nonlinear_part_simplification_4}. 
	\end{subequations}
	Substituting \cref{seq:ampl_eqn_nonlinear_part_simplification_4} back into the amplitude equation, we get
	\begin{align}\label{seq:amplitude_equation_expanded}
		\left(w_2k_s^2\left(\frac{\sin\left(\frac{n\pi}{2}\right)}{n\pi}-\frac{1}{4}\right)-w_0 \right) a_n + w_2k_s^2\frac{\sin\left(\frac{n\pi}{2}\right)}{2\pi} \left(\sum_{q=1}^{n-1}\frac{(-1 + (-1)^q)a_{n-q}a_q}{n-2q}+(1 - (-1)^n)\sum_{q=1}^{\infty}\frac{(-1)^q a_{n+q}a_q}{n+2q}\right)=0,
	\end{align}
	for steady state. Since $\sin\left(\frac{n\pi}{2}\right)=0$ for even $n$, the nonlinear term is zero for even $n$ unless $n=2q$. This leads to
	\begin{equation}
		a_{2q}=\frac{a_q^2}{4\frac{w_0}{w_2k_s^2}+1}.
	\end{equation}
	Truncating \cref{seq:amplitude_equation_expanded} up to the second mode, we get a quadratic equation for the order parameter, which has the solution
	\begin{equation}
		a_1\approx\frac{1}{4k_s}\left(\frac{3(1+4w_0)}{\pi w_2}(w_0^{\mathrm{cr}}-w_0)\right)^{\frac{1}{2}}.
	\end{equation}
	Truncating \cref{seq:amplitude_equation_expanded} up to 4th order leads to the best possible analytical approximation, that can be obtained through this method. The resulting equation for $a_1$ was obtained using the symbolic language \textsc{Mathematica}~\cite{Mathematica} and is too large to reasonably be displayed in this supplement. The plot of the 4th order approximation of $a_1$ is displayed as a solid black line in figure 3 of the main text. Truncating at 5th order already leads to a generic quintic polynomial for $a_1$, which cannot be solved analytically~\cite{rosen1995niels}.
	
	\section{Phase transition in the MCA model}\label{ssec:mca_derivations}
	Since our numerical data shows that the phase transition in the MCA model is first order, we expect to have two phase boundaries. One can be obtained from the stability analysis of the ordered phase and the other from the stability analysis of the disordered phase. Using the methods described in \cref{{ssec:mf_derivations}}, we could only perform the stability analysis of the disordered phase. Analyzing the stability of the ordered phase would involve expanding the state density $f(\theta;t)$ around $\delta(\theta)$, which is was not feasible. 
	\subsection{Stability analysis of the disordered phase in MCA model}
\begin{figure}[h]
		\centering
		\includegraphics{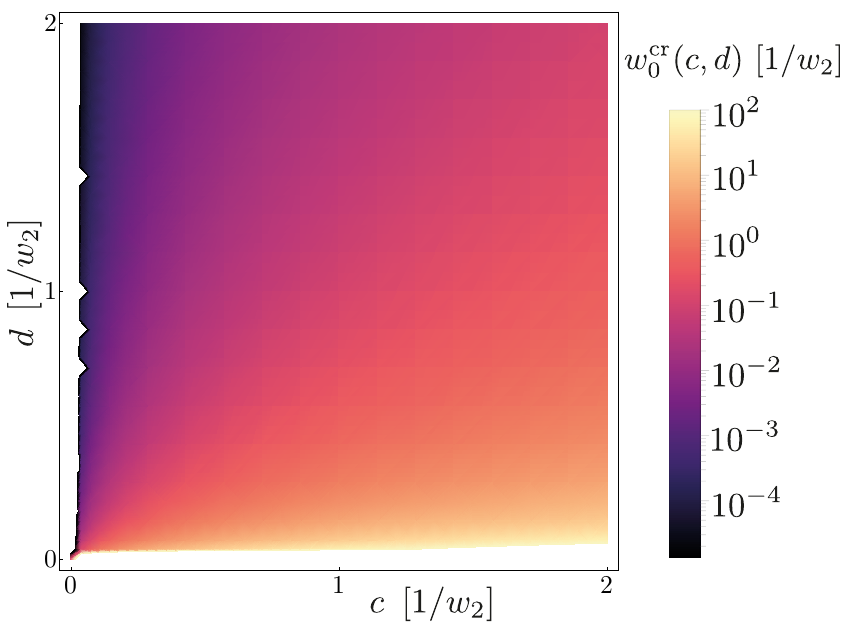}
		\caption{Critical surface $w_0^{\mathrm{cr}}(c,d)$. For values below $w_0^{\mathrm{cr}}(c,d)$,  disordered phase is linearly unstable.}
		\label{fig:supplement_figure}
	\end{figure}

	Again we assume the symmetry of $f(\theta;t)$, furthermore we assume that, at least to the linear order in Fourier modes, we can expand $l_2(\theta,\theta';t)$ separately in its arguments, i.e.
	\begin{align}
		l_2(\theta,\theta';t)&=\frac{k_s}{4\pi^2}\bigg(1+2\sum_{n=1}^{\infty}b_n \cos(n\theta) + 2\sum_{n=1}^{\infty}b_n \cos(n\theta')\bigg) + \mathcal{O}(b_n b_m)\\
		&=\frac{k_s}{4\pi^2}\bigg(1+2\sum_{n=1}^{\infty}b_n \left[\cos(n\theta) +\cos(n\theta')\right]\bigg),
	\end{align}
	where we also dropped the anti-symmetric part of the Fourier expansion. We are only interested in stability of the state density distribution, which we assume to be symmetric, and since to the linear order the time evolution of symmetric and anti-symmetric modes decouples, we make the analysis easier, by only considering the symmetric modes of the link density distribution. Note that to the linear order, there is no additional information stored in the second variable of $l_2$, this means that we can integrate it out to simplify \cref{seq:mca_system}, which gives us
	\begin{subequations}\label{seq:mca_marginalzed}
		\begin{align}
			\partial_t f(\theta;t)=&w_0\left(\frac{1}{2\pi}-f(\theta;t)\right)+\frac{k_s^2}{4}\bigg[\int_{-\pi}^{\pi}\mathrm{d}\theta'\frac{I_{\frac{1}{2},l}(\theta';\theta;t)}{f(\theta';t)}-\frac{l^2(\theta;t)}{f(\theta;t)}\bigg]\\
			\partial_t l(\theta;t)=&w_0\left(\frac{1}{2\pi}-l(\theta;t)\right)+d\big(f(\theta;t)-l(\theta;t)\big) 
			+k_s\bigg[\int_{-\pi}^{\pi}\mathrm{d}\theta'\frac{I_{1,l}(\theta';\theta;t)}{f(\theta';t)}-\frac{l^2(\theta;t)}{f(\theta;t)}\bigg] \nonumber \\
			&+\frac{k_s^2}{4}\bigg[\int_{-\pi}^{\pi}\mathrm{d}\theta'l(\theta')\frac{I_{1,l}(\theta';\theta;t)}{f^2(\theta';t)}-\frac{l^3(\theta;t)}{f^2(\theta;t)}\bigg],
		\end{align}
	\end{subequations}
	where, we already set $w_2=1$ which rescales time and 
	\begin{subequations}\label{seq:mca_aux_def}
		\begin{align}
			l(\theta;t)=&\int_{-\pi}^{\pi}\mathrm{d}\theta'\; l_2(\theta,\theta';t)=\frac{1}{2\pi}\bigg(1+2\sum_{n=1}^{\infty}b_n \cos(n\theta)\bigg) +\mathcal{O}(b_n b_m) \label{seq:l_fourier_expansion}, \\
			I_{\frac{1}{2},l}(\theta;\theta';t)=&\int_{-\pi}^{\pi}\mathrm{d}\xi\; l_2(\theta,\theta'+\frac{\xi}{2};t)\; l_2(\theta,\theta'-\frac{\xi}{2};t), \\
			I_{1,l}(\theta;\theta';t)=&\int_{-\pi}^{\pi}\mathrm{d}\xi\; l_2(\theta,\theta'+\xi;t)\; l_2(\theta,\theta'-\xi;t).
		\end{align}
	\end{subequations}
	We plugged the Fourier expansions of \cref{seq:fourier_expansion} and \cref{seq:l_fourier_expansion} in \cref{seq:mca_marginalzed}, and kept only linear terms in Fourier modes. Since \cref{seq:mca_marginalzed} consists of much longer equations then \cref{seq:mf_short}, we automated this process with \textsc{Mathematica}~\cite{Mathematica} and got,
	\begin{equation}
		\partial_t
		\begin{pmatrix}
			a_n \\ 
			b_n
		\end{pmatrix}
		=
		\boldsymbol{M}
		\cdot
		\begin{pmatrix}
			a_n \\ 
			b_n
		\end{pmatrix},
	\end{equation}
	where 
	\begin{align}
		\boldsymbol{M} = \left(
		\begin{array}{cc}
			\frac{c^2-4 d^2 \text{w0}}{4 \pi  d^2} & -\frac{c^2 \left(\pi  q-2 \sin \left(\frac{\pi 
					q}{2}\right)\right)}{2 \pi ^2 d^2 q} \\
			\frac{c^2+2 c d+2 d^3}{2 \pi  d^2} & \frac{4 c^2 \sin \left(\frac{\pi  q}{2}\right)-\pi  q \left(3
				c^2+8 c d+4 d^2 (d+\text{w0})\right)}{4 \pi ^2 d^2 q}. \\
		\end{array}
		\right)
	\end{align}
	In this case the eigenvalues of $\boldsymbol{M}$ indicate the stability of the system. Only the largest eigenvalue can become positive, which gives us a condition 
	\begin{equation}\label{seq:w_0cd}
		w_0^{\mathrm{cr}}(c,d)=\frac{-\pi  \left(c^2+4 c d+2 d^3\right)+2 c^2+2 \sqrt{c^4+2 \pi  c^2 d^3+\pi ^2 d \left(2 c^3+4 c^2
				d+4 c d^3+d^5\right)}}{4 \pi  d^2},
	\end{equation}
	for critical noise. Below $w_0^\mathrm{cr}$ the disordered phase is linearly unstable. We plot the critical surface $w_0^{\mathrm{cr}}(c,d)$ in \cref{fig:supplement_figure}. We see that the critical surface trivially depends on $c$ and $d$. For $c>>d$ the values of $w_0^{\mathrm{cr}}(c,d)$ become very large, which is expected since $c>>d$ is equivalent to a very dense network. In order to stay in the sparse network limit and simplify the expressions, we set $c=d$, which is the convention that we used in the main text. Making this substitution in \cref{seq:w_0cd} and solving for $d$ leads us to eq.~(10) from the main text.
	\section{Integral transformations}\label{ssec:integral_transformations}
	Here we show two general results that we used to arrive at simplified forms of differential equations.
	First we used the sign reversal symmetry of the $\xi$ integral
	\begin{equation}\label{seq:xi_symmetry}
		\int_0^{\frac{\pi}{2}} \mathrm{d}\xi\; g(\theta-\xi) h(\theta+\xi) 
		= \frac{1}{2} \int_{-\frac{\pi}{2}}^\frac{\pi}{2} \mathrm{d}\xi\; g(\theta-\xi) h(\theta+\xi) 
		= \frac{1}{4} \int_{-\pi}^\pi \mathrm{d}\xi\; g\left(\theta-\frac{\xi}{2}\right) h\left(\theta + \frac{\xi}{2}\right),
	\end{equation}
	and after defining
	\begin{equation}
		I(\theta) 
		= \int_{-\pi}^\pi \mathrm{d}\xi\; g\left(\theta-\frac{\xi}{2}\right) h\left(\theta + \frac{\xi}{2}\right),
	\end{equation}
	we show a general relation,
	\begin{subequations}\label{seq:integralis_egregium_generalis}
		\begin{align}
			\int_{-\pi}^{\pi} \mathrm{d}\theta I(\theta)
			=& \hspace*{1.5em} \int_{-\pi}^{\pi} \mathrm{d}\theta\; \int_{-\pi}^{\pi} \mathrm{d}\xi\; g\left(\theta-\frac{\xi}{2}\right) h\left(\theta + \frac{\xi}{2}\right) \\
			\stackrel{\mathclap{\theta\to\theta+\xi/2}}{=}& \hspace*{1.5em} \int_{-\pi}^{\pi} \mathrm{d}\xi\;  \int_{-\pi-\frac{\xi}{2}}^{\pi-\frac{\xi}{2}} \mathrm{d}\theta\; g(\theta)h(\theta+\xi) \\
			=& \hspace*{1.5em} \int_{-\pi}^{\pi} \mathrm{d}\xi\;  \int_{-\pi}^{\pi} \mathrm{d}\theta\; g(\theta)h(\theta+\xi)\label{seq:backshift_theta} \\
			\stackrel{\mathclap{\xi\to\xi-\theta}}{=}& \hspace*{1.5em} \int_{-\pi}^{\pi} \mathrm{d}\theta\; g(\theta) \int_{-\pi+\theta}^{\pi+\theta} \mathrm{d}\xi\; h(\xi) \\
			=& \hspace*{1.5em} \int_{-\pi}^{\pi} \mathrm{d}\theta\;  g(\theta) \int_{-\pi}^{\pi} \mathrm{d}\xi\; h(\xi) \label{seq:backshift_xi} \\
			= & \hspace*{1.5em}G H &
		\end{align}   
	\end{subequations}
	which holds for any functions $g(\theta)$ and $h(\theta)$ which are periodic on $(-\pi,\pi]$. We used the property of periodicity in \cref{seq:backshift_theta} and \cref{seq:backshift_xi} to discard the shift of the integral boundaries, since the integral is over the whole domain of a periodic function. And $G$ and $H$ denote the integrals over the full period of the functions $g$ and $h$ respectively.
	%